\documentclass[traditabstract]{aa}
\usepackage{graphicx,natbib,xcolor,amsmath}
\usepackage{txfonts,afterpage}

\def\lta{ \lower .75ex\hbox{$\sim$} \llap{\raise .27ex \hbox{$<$}} }

\begin{document}

\date{Received .../Accepted ...}

\title{Radiation spectra of warm and optically thick coronae in AGN}
\author{P.-O. Petrucci\inst{1}
  \and D. Gronkiewicz\inst{2}
  \and A. Rozanska\inst{2}
  \and R. Belmont\inst{3}
  \and S. Bianchi\inst{4}
  \and B. Czerny\inst{5}
 \and G. Matt\inst{4}
  \and J. Malzac\inst{6}
 \and R. Middei\inst{4}
 \and A. De Rosa\inst{7}
  \and F. Ursini\inst{8}
 \and M. Cappi\inst{8}
  }
  
\institute{Univ. Grenoble Alpes, CNRS, IPAG, F-38000 Grenoble, France 
	   \and
	   Nicolaus Copernicus Astronomical Center, Polish Academy of Sciences, Bartycka 18, 00-716, Warsaw, Poland
	\and
	   Laboratoire AIM (CEA/IRFU - CNRS/INSU - Universit?e Paris Diderot), CEA DSM/IRFU/SAp, F-91191 Gif-sur-Yvette, France
	   \and
	   Dipartimento di Matematica e Fisica, Universit\`a degli Studi Roma Tre, via della Vasca Navale 84, I-00146 Roma,Italy.
	   \and
	   Center for Theoretical Physics, Polish Academy of Sciences, Al. Lotnikow 32/46, 02-668 Warsaw, Poland
	   \and 
	   IRAP, Universite de Toulouse, CNRS, UPS, CNES, Toulouse, France
	   \and 
	   INAF-Istituto di Astrofisica e Planetologie Spaziali, Via Fosso del Cavaliere, 00133 Rome, Italy
	   \and 
	   INAF-Osservatorio di astrofisica e scienza dello spazio di Bologna (OAS), Via Piero Gobetti 93/3, I-40129 Bologna, Italy
} 

%

\abstract{A soft X-ray excess above the 2-10 keV power law extrapolation is generally observed in AGN X-ray spectra. The origin of this  excess is still not well understood. Presently there are two competitive models: blurred ionized reflection and warm Comptonisation. In the case of warm Comptonisation, observations suggest a corona temperature in the range 0.1-2 keV and a corona optical depth $\sim$10-20. Moreover, radiative constraints from spectral fits with Comptonisation models suggest that most of the accretion power should be released in the warm corona and the disk below is basically non-dissipative, radiating only the reprocessed emission from the corona. The true radiative properties of such a warm and optically thick plasma are not well-known, however. For instance, the importance of the Comptonisation process, the potential presence of strong absorption/emission features or the spectral shape of the output spectrum have been studied only very recently. We present in this paper simulations of  warm and optically thick coronae using the {\sc titan} radiative transfer code coupled with the {\sc noar} Monte-Carlo code, the latter fully accounting for Compton scattering of continuum and lines. Illumination from above by a hard X-ray emission and from below by an optically thick accretion disk is taken into account as well as (uniform) internal heating.  
Our simulations show that for a large part of the parameter space, the warm corona with sufficient internal mechanical heating is dominated by Compton cooling and neither strong absorption nor emission lines are present in the outgoing spectra. In a smaller part of the parameter space, the calculated emission agrees with the spectral shape of the observed soft X-ray excess. Remarkably, this also corresponds to the conditions of radiative equilibrium of an extended warm corona covering almost entirely a non-dissipative accretion disk. These results confirm the warm Comptonisation as a valuable model that can explain the origin of the soft X-ray excess.}


\maketitle


\section{Introduction}
When extrapolating the 2-10 keV power law of type 1 (unabsorbed) Active Galactic Nuclei (AGN) down to soft X-rays ($<$ 2 keV), most of the objects show an excess in emission, the so-called soft X-ray excess. This excess is seen in a large majority of AGN \citep[e.g.][]{wal93,pag04,gie04,bia09a} and its origin is still not well understood. Two interpretations are generally discussed: blurred ionized reflection (e.g. \citealt{cru06,wal13}) or Comptonisation in a warm ($T\sim$ 1 keV) and optically thick ($\tau\sim$10-20) corona \citep[e.g.][hereafter P18, and references therein]{mag98,jin12a, pet13,por18,pet18}. \\ 

Both interpretations give a good fit to the data but both have their own limitations (see discussion in \citealt{gar18}). Blurred ionized reflection models generally require extreme values for the spin and hot corona compactness, as well as large ionisation degree and large densities (see e.g. \citealt{jia19b,jia19a}) which can also significantly reduce the inferred iron abundances compared to typical (e.g. with density generally equal to $10^{15}$ cm$^{-3}$) reflection models (e.g. \citealt{tom18}). On the other hand, only a few attempts of realistic modelings of the warm corona emission have been done so far (e.g. \citealt{bal20}). At such low temperature (around 1 keV) large atomic opacities could dominate over the Thomson opacities. This would have two major consequences: Comptonisation would not dominate the emitting processes and strong absorption and/or emission features could be present in the output spectrum in contradiction with observations. If true, this would invalidate warm Comptonisation as an origin of the soft X-ray excess.\\
\begin{figure}[t]
\begin{center}
\includegraphics[width=\columnwidth]{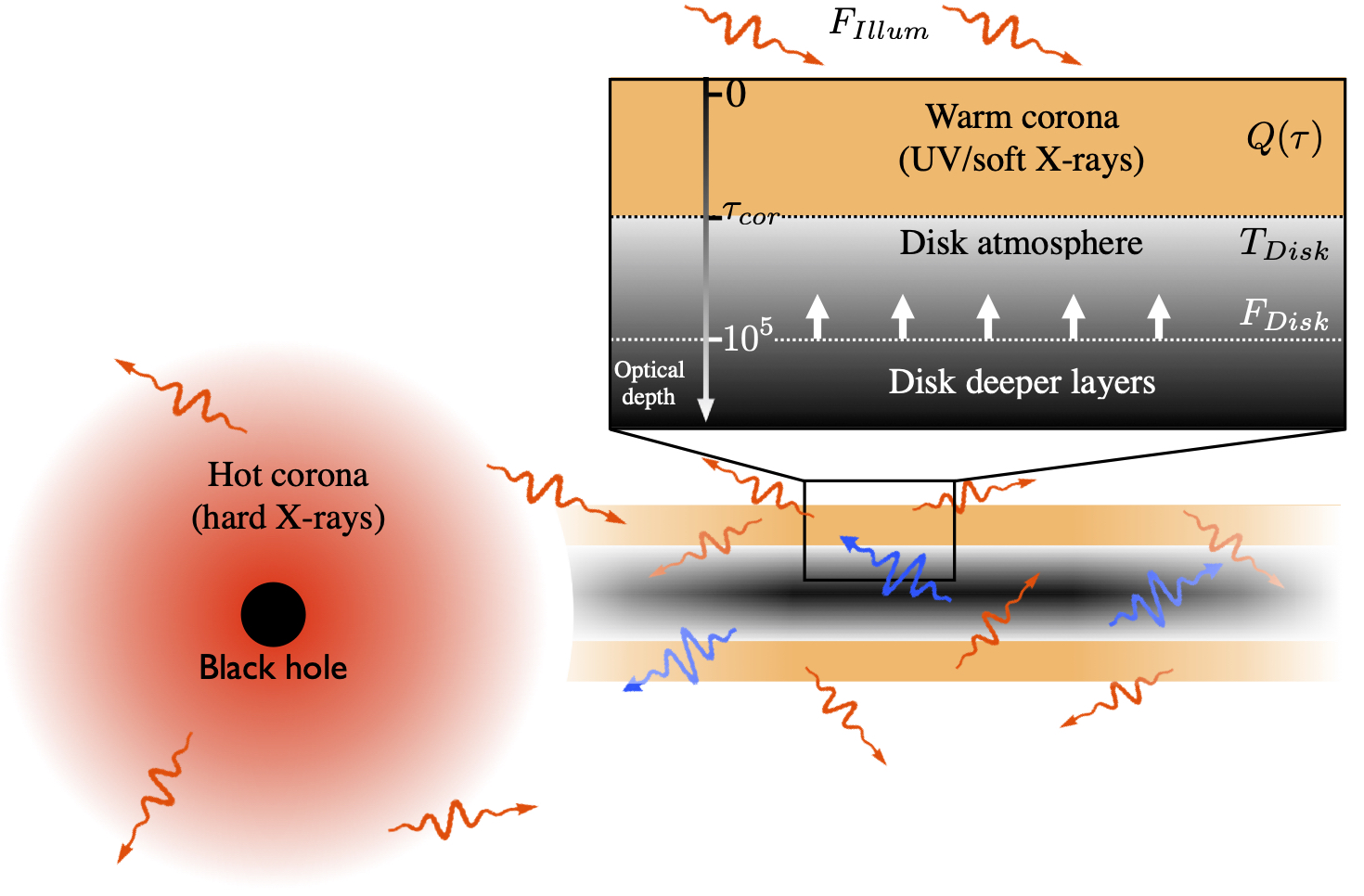}
\caption{Sketch of the two coronae approach. A hot ($kT\sim$ 100 keV) and optically thin ($\tau\sim$1) corona (in red) is present in the inner parts of the accretion flow, close to the black hole. It produces the hard X-ray emission through Comptonisation of the soft photons coming from the outer part of the flow. The outer accretion flow is a vertically structured accretion disk (a zoom is shown in the figure) with cold and optically thick matter in the deeper layers (in black), characterized by a tempertature $T_{\rm bb}$ and threaded by a vertical radiative flux $F_{\rm disk}$, while the upper layers (in orange) are composed by a warm ($kT\sim$ 1 keV) and optically thick ($\tau_{\rm cor}\sim$10-20) corona possessing a source of internal heating power $Q(\tau)$. The warm corona is also illuminated from above by an illumination flux $F_{\rm Illum}$ produced by the hot corona.This sketch is partly inspired from \cite{roz15}.}
\label{FigSketch}
\end{center}
\end{figure}

The warm Comptonisation modeling of the soft X-ray excess was carefully tested with the data set from the large broad band campaign on Mrk 509 \cite[][hereafter P13]{kaa11a,pet13}. The data were fitted with the so-called ``two-coronae'' scenario (see the sketch of Fig. \ref{FigSketch}). It assumes two Comptonisation model components, one for the UV-Soft X-rays, to simulate the warm corona emission, and one for the hard X-rays, to simulate the standard hot corona emission. The model gave statistically good fits for all the observations of the campaign and it provided very interesting information on the geometry of each corona (see P13 for more details). Importantly for the present paper, the warm corona was found to agree very well with a powerful, extended and optically thick plasma covering a non-dissipative accretion disk, i.e., all the accretion power would be released in the warm corona, the underlying disk radiating only the reprocessed emission produced by the warm corona. Analyses of larger AGN samples give similar conclusions (e.g. P18). While the "two-coronae" approach has been successfully applied to several AGNs (e.g.\citealt{mag98,cze03,jin12a,mat14,meh15,urs16,urs18,mid18a,por18,kub18,mid19}) fundamental questions remain unanswered: can a warm and optically thick corona exist in the upper layers of the accretion flow? Can its spectral emission be dominated by Comptonisation and be in agreement with the observed soft and smooth X-ray spectral shape? \\

A realistic radiative transfer study of an optically thick and warm plasma is not simple however. The optical depth is neither very small nor very large so that radiative transfer cannot be simply approximated. As previously said, the temperature is also in a range of values where lots of ions are expected and line emission/absorption processes have to be included.  Last but not least, the corona heating power has to be locally released as suggested by observations. \citet[][hereafter R15]{roz15} did a theoretical study of the existence of such a warm optically thick corona at the surface of a standard accretion disk assuming both radiative and hydrostatic equilibrium. This study shows that it is  indeed possible to obtain solutions having the required temperature. Specifically the best conditions are obtained when most of the power of accretion is released into the corona rather than the disk. Nevertheless, the requirement of hydrostatic equilibrium with the underlying disk puts an upper limit of $\sim$5 on the warm corona optical depth. However R15 also showed that for larger optical depth, large magnetic to gas pressure is needed ($>$30) to keep such a thick corona in hydrostatic equilibrium with negligible free-free emission. The impact of magnetic pressure on the formation of the warm corona has been recently studied in the case of accretion disks in X-ray binaries (XrB) by \cite{gron19}. While the estimates of R15 suggest that a warm corona could indeed exist, no spectral emission was simulated. Line radiative transfer was also not included in the computation preventing a clear answer to the limitations associated with this model.\\

The goal of this paper is to simulate energy dependent spectra of warm coronae in radiative equilibrium thanks to the state-of-the-art radiative transfer code {\sc titan} coupled with the Monte-Carlo code {\sc noar} \citep{dum03}. We first introduce a few parameters and recall in Sect. \ref{transrad} some analytical estimates of radiative transfer equations in the case of a grey optically thick scattering medium with dissipation. We detail the main characteristics of the {\sc titan} and {\sc noar} codes in Sect. \ref{titancode} and present the results of our simulations in Sect. \ref{simu}. We discuss these results in Sect. \ref{summary} before concluding.

\section{Analytical estimates}
\label{transrad}
We recall here the analytical solutions of radiative transfer equations for the warm corona with additional heating located above an accretion disk. To ease the analysis, we adopt here a number of simplifying assumptions with respect to the numerical
calculations described in the next section. Actually, here we assume a grey atmosphere with pure scattering (i.e. we neglect emission and absorption from bound-free and bound-bound transitions). We also assume that the atmosphere is optically thick, and therefore adopt the Eddington approximation in the whole medium. We assume a constant density $n_{\rm H}$ throughout the atmosphere and we call $\tau_{\rm cor}$ its total Thomson optical depth. The corona is assumed to be heated with a uniform rate per unit optical depth and per unit solid angle $Q$ (erg s$^{-1}$ cm$^{-2}$ str$^{-1}$). For simplicity, we neglect any illumination from above the corona. The  disk is supposed to radiate like a black-body of temperature $T_{\rm bb}$ and we call $B$ the frequency integrated black-body radiation intensity, i.e. $B(T_{\rm bb})=\sigma T_{\rm bb}^4/\pi$. The disk is also characterised by its intrinsic flux $F^{\rm int}_{\rm disk}$ produced  through internal dissipation (see Fig. \ref{FigSketch}). The disk can still radiate ($B>$0) even if the disk is non-dissipative i.e. $F^{\rm int}_{\rm disk}=0$. In this case it radiates only by reprocessing the corona flux emitted downward towards the disk. On the other hand, if the corona intercepts only a small fraction of the disk emission, $B$ can be very small, and potentially vanish. \\

We follow the analytical computations of R15, but we present the equations in a slightly different way which will be useful to understand the results of the numerical simulations presented in the next sections. 
From now on, the part of the radiative transfer variables emitted upward with respect to the disk will have a ``$^+$'' exponent and the part emitted downward will have a ``$^-$''.\\

The frequency-integrated radiation transfer equation with an additional input energy rate per unit
optical depth and solid angle $Q$, can be written as:
\begin{equation}
\mu\frac{dI}{d\tau}=I-J-Q \label{eqtrans}
\end{equation}
where $\mu$ is the cosine of the azimuthal angle. The optical depth, $\tau$ is measured downward, from the top of the corona toward the disk, $I(\mu,\tau)$ is the radiation true intensity, and $J(\tau)$ is the radiation mean intensity, i.e. {averaged over the angles}. \\
In a grey atmosphere and assuming the Eddington approximation in the whole medium, the parts of the radiation intensity emitted upward ($\mu>$0) $I^+(\tau)$ and backward ($\mu<$0) $I^-(\tau)$  are given by:
\begin{eqnarray}
I^+(\tau) &=& J(\tau)+2H(\tau)\label{Ip} \\
I^-(\tau) &=& J(\tau)-2H(\tau) 
\end{eqnarray}
where $H(\tau)$ is the zeroth moment of Eq. \ref{eqtrans}. It gives the net flux crossing the corona at each optical depth $\tau$ inside the slab. Following R15, the expressions of $J(\tau)$ and $H(\tau)$ are given by:
\begin{eqnarray}
J(\tau) &=& 3\frac{F_{\rm out}}{4\pi}\left [ \frac{2}{3}+\tau-\frac{\chi \tau^2}{2\tau_{\rm cor}}\right ] \label{eqJ}\\
H(\tau) &=& \frac{F_{\rm out}}{4\pi}\left (1-\frac{\chi\tau}{\tau_{\rm cor}}\right )\label{eqH}
\end{eqnarray}
where $F_{\rm out}$ is the corona escaping radiation flux at its surface\footnote{$F_{\rm out}$ is noted $F^{\rm tot}_{\rm acc}$ in R15} i.e. at $\tau$=0. In these equations, the parameter $\chi$ is defined by 
\begin{equation}
\chi =\frac{F_{\rm cor}}{F_{\rm out}}
\label{eqChi}
\end{equation}
where $F_{\rm cor}$ is the flux produced through dissipation in the corona: 
\begin{equation}
F_{\rm cor} = 4\pi Q\tau_{\rm cor}.\label{Fcor}
\end{equation}
From Eq. \ref{eqH}, $\chi$ can be rewritten as:
\begin{equation}
\chi =\frac{\tau_{\rm cor}Q}{H({\tau_{\rm cor}})+\tau_{\rm cor}Q}\label{eqChiH}
\end{equation}
Now, at the base of the corona, we simply have:
\begin{equation}
I^+(\tau_{\rm cor})=J(\tau_{\rm cor})+2H(\tau_{\rm cor})=B\label{eqI+B}
\end{equation}

Combining Eqs. \ref{eqJ}, \ref{eqH} and  \ref{eqI+B}, we can express 
$H(\tau_{\rm cor})$ in function of the different parameters of the problem:
\begin{equation}
H(\tau_{\rm cor})=\frac{B}{4+3\tau_{\rm cor}}-\frac{\tau_{\rm cor}Q}{2}\label{eqHgen}
\end{equation}
The function $H(\tau_{\rm cor})$ gives the net flux crossing the base of the corona i.e. the difference between the flux emitted upward inside the corona and the flux emitted downward by the corona inside the disk. This difference cannot be smaller than $-\tau_{\rm cor}Q/2$ which is reached in the extreme case of no disk emission ($B$=0). \\

The value of $H(\tau_{\rm cor})$ can be interpreted in terms of different conditions of the disk-corona radiative equilibrium:
\begin{itemize}
\item $H(\tau_{\rm cor})=0$: this corresponds to $\chi=1$. This is the condition when the upward flux compensate exactly the downward one at the base of the corona. In the case of a warm corona covering entirely the disk, this imposes that the disk is non-dissipative i.e. $F_{\rm disk}^{\rm int}$=0
\item $H(\tau_{\rm cor})>0$: this corresponds to $\chi<1$. In this case, the upward flux entering the warm corona at its base is larger than the downward one. This is for example the case of slab corona above a dissipative disk with $F_{\rm disk}^{\rm int}>$0
\item $H(\tau_{\rm cor})<0$ but $>-\frac{\tau_{\rm cor}Q}{2}$: this corresponds to $2>\chi>1$. In this case, the upward flux entering the warm corona at its base is smaller than the downward one. This is for example the case of a patchy corona above a non-dissipative disk ($F_{\rm disk}^{\rm int}$=0), where part of the disk emission does not reenter entirely in the warm corona.
\end{itemize}
These analytical results do not take into account the heating due to external illumination nor other radiation processes (lines, edges, free-free). The full treatment of the radiation transfer problem requires numerical simulations presented in the following sections.

\begin{figure*}[t]
\begin{center}
\begin{tabular}{cc}
\includegraphics[height=0.5\textwidth]{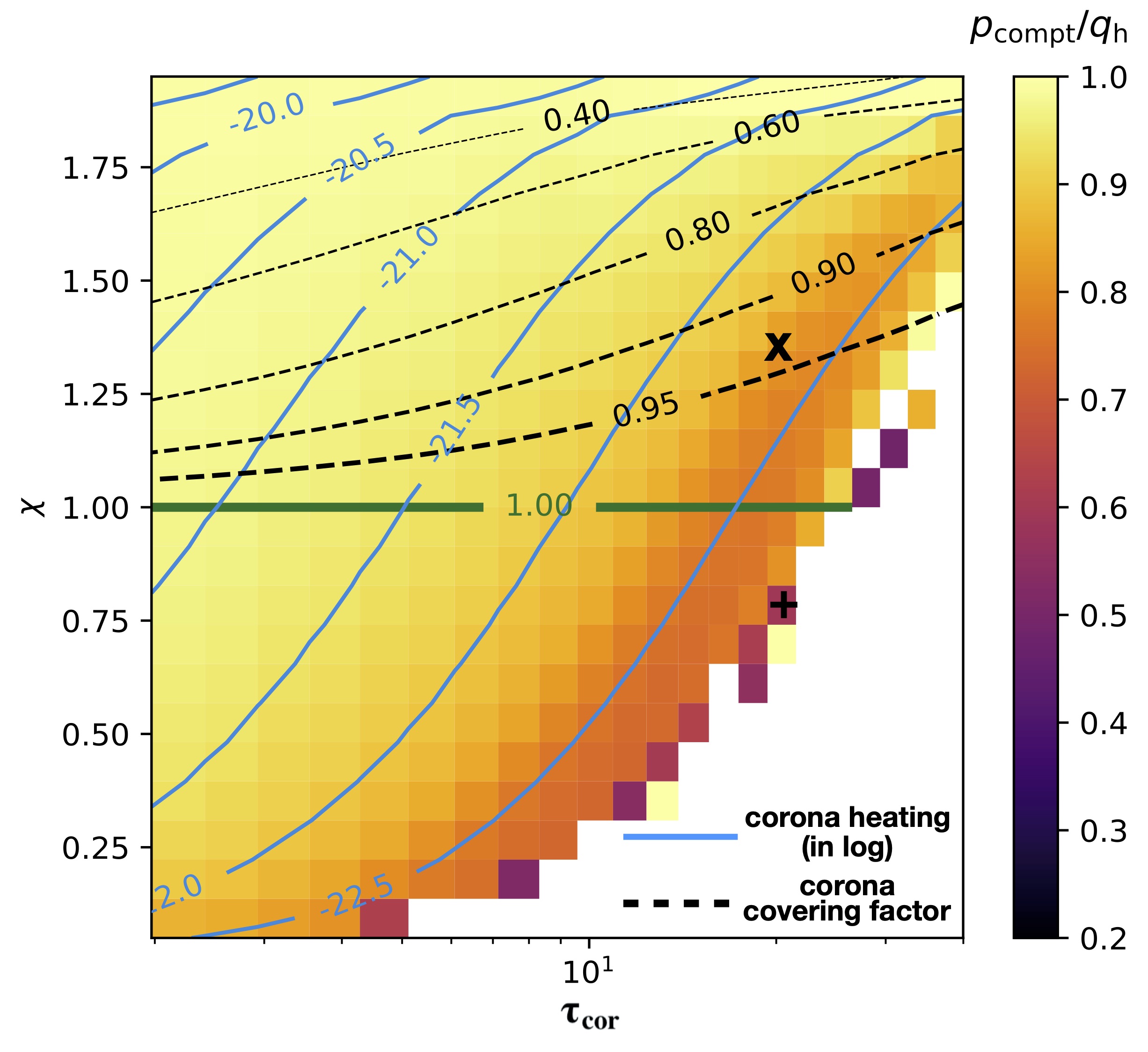} &
\includegraphics[height=0.48\textwidth,width=0.4\textwidth]{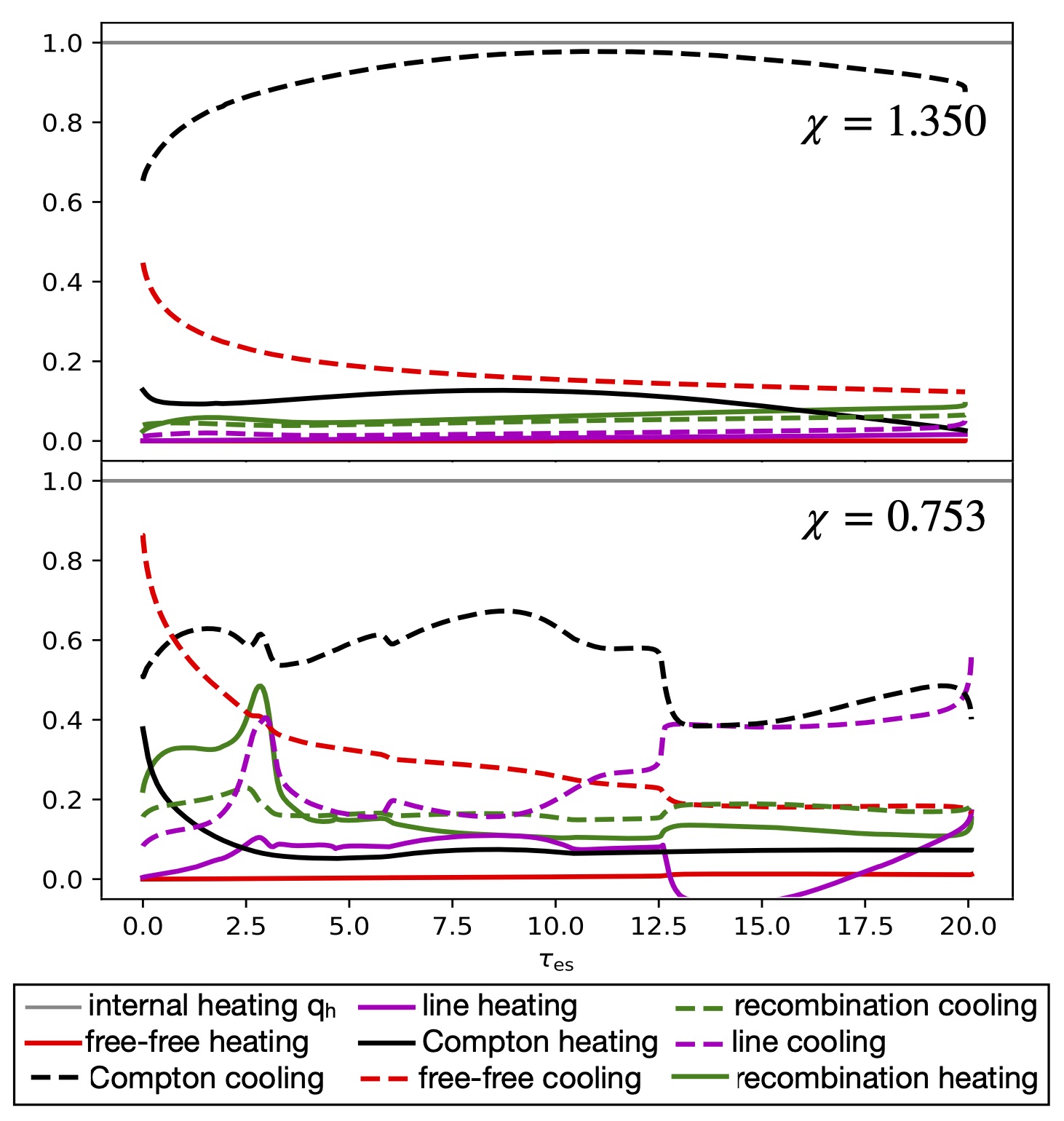}
\end{tabular}
 \caption{{\bf Left:} Map of the average Compton cooling $\displaystyle \left (p_{\rm compt}=\frac{1}{\tau_{cor}}\int_{0}^{\tau_{\rm cor}} p_{\rm compt}(\tau)d\tau \right )$ to the warm corona heating ($q_{\rm h}$) ratio in the $\chi$-$\tau_{\rm cor}$ plane, $\chi$ being defined by Eq. \ref{eqChisimu}. The color scale, on the right, ranges from 20 (black) to 100\% (white). The other parameters are fixed to $T_{\rm bb}=10^5$ K, {\boldmath$\xi_0=10^3$} erg s$^{-1}$ cm, and $n_{\rm H}=10^{12}$ cm$^{-3}$.  The solid green line refers to $\chi$=1, i.e. the case of radiative equilibrium between a corona covering entirely a non-dissipative disk. It divides the $\chi$-$\tau_{\rm cor}$ in two regions that can be characterized by different radiative equilibrium properties (see Sect. \ref{transrad}). The region above the green line, i.e. $\chi>1$, agrees with a non-dissipative disk covered by a patchy corona. 
 The dashed lines correspond to different values of the corona covering factor from 0.95 to 0.40 {(see Eq. \ref{eqChiA} in Appendix)}.  The region below the green line, i.e. $\chi<1$, agrees with a slab corona  above a disk which is now dissipative ($F_{\rm disk}>$0). The blue contours correspond to different values of the logarithm of the warm corona heating $q_{\rm h}$ (in unit of erg s$^{-1}$ cm$^{3}$). {\bf Right}: Cooling and heating process fractions with respect to the warm corona heating $q_{\rm h}$ across the corona for $\tau_{\rm cor}=20$ but for two different values of $\chi$, 0.753 (bottom) and 1.35 (top). This correspond to the simulations indicated with a "+" and "x" respectively on the left figure. }
\label{FigSimu1ab}
\end{center}
\end{figure*}

\section{Simulations}
\label{simu}
\subsection{TITAN and NOAR code}
\label{titancode}
To calculate the radiative equilibrium and the output spectra of a dissipative warm corona we have used the newest version of  
the radiative transfer code {\sc titan}  \citep{dum03} coupled with the Monte-Carlo {\sc noar} code \citep{dum00,noar2001}


{\sc titan} is a one-dimensional code that allows to compute the structure and angle dependent emergent spectra of
hot photoionized gas by solving the radiative transfer simultaneously with ionization and thermal equilibrium. This can be done assuming a constant density, a constant pressure or an hydrostatic equilibrium prescription.
On the ``top'' side of the gas slab, {\sc titan} allows the illumination by radiation of any spectral energy
distribution (SED) {and with any angular dependence}, and on the ``disk'' side it allows to set the boundary condition for black-body illumination
to account for the presence of a colder disk underneath.

Assuming the balance between ionization and recombination of ions, excitation and de-excitation,
and most important atomic lines, {\sc titan} computes a physical state of the gas at each depth in thermal and pressure
equilibirum. About 4000 line transitions are included in {\sc titan} code.  
The population of each ion level is computed by solving the
set of ionization equations coupled with the set of statistical
equations describing the excitation equilibrium in full non-LTE (non local thermal equilibrium) approach.
Free-free emission is fully taken into account. The code was designed to work with optically thick media and solves the radiative transfer equation both in
continuum and lines using Accelerated Lambda Iteration (ALI) method \citep{collin2004}. The computations
are done assuming complete redistribution function in
the lines. Partial redistribution is mimicked by a Doppler profile
for some of the most intense resonant lines.

{The Compton heating/cooling balance of the matter is taken into account by solving simultaneously for the temperature structure and the radiation field. Although Titan includes most relevant processes, it does not take into account the Compton heating of the radiation field and the corresponding photon energy shift via Comptonization in the emergent spectra. The former effect is mainly significant for $\chi >1$  and, in this case, the temperature computed by  {\sc titan} can be over estimated by a factor of a few. To take into account the latter effect, we use the Monte Carlo code {\sc noar} that takes the density, temperature and ionization structure of the slab calculated by  {\sc titan} and computes the emergent spectra on both sides of the slab. }

The coupling between {\sc titan} and {\sc noar} allows a complete treatment of the emission from a photoionized, Comptonized medium.
It can be used in a variety of cases as: illuminated disk atmospheres \citep{rozanska2002}
and warm absorbers \citep{rozanska2006}  in AGN. \\

For the purpose of this paper we consider a constant density warm corona {and an illumination from above normal to the disk\footnote{{We do not expect the effects of anisotropy of the illumination to have a significant influence on the thermal structure of the warm corona (which is the main goal of this publication). Indeed, after a few $\tau$, these anisotropy effects will be spread-out by the multiple photon scatterings.}}}. The computational pipeline that we built runs {\sc titan} with different values of the following parameters: gas density $n_{\rm H}$ {in the range [$10^9$--$10^{14}$] cm$^{-3}$}, total optical depth
$\tau_{cor}$ {in the range  [2--40]}, local heating rate $q_{\rm h}$ {in the range [{\boldmath$10^{-23}$--$10^{-19}$}] erg s$^{-1}$ cm$^{3}$}, ionization parameter {\boldmath$\xi_0$} of the ``top" side illumination\footnote{{By definition $\xi_0=F_{illum}/n_{\rm H}$ with $F_{illum}$ the illuminating flux impinging on the slab surface (see Fig. 1)}}  (assuming a power law with spectral photon index of 1.8 in the energy range 50 eV-100 keV) {in the range [10--3.10$^{5}]$ erg s$^{-1}$ cm} and illumination black-body temperature $T_{\rm bb}$ on the ``disk'' side {in the range [3.10$^{4}$--10$^{6}$] K}.
{We emphasize that the ionisation parameter {\boldmath$\xi_0$} is directly linked to the flux of the incoming radiation at the surface of the slab, but {\sc titan} takes also into account the spectral shape of this incoming radiation in the radiative transfer computation.}
The mechanical heating rate 
$q_{\rm h}$ is supposed to be uniform
throughout the entire slab. It is related to the heating power per unit optical depth and solid angle $Q$ defined in Sect. \ref{transrad}:
\begin{equation}
q_{\rm h}=4\pi Q \sigma_{\rm T}/n_{\rm H}\label{eqqh}.
\end{equation}

{\sc titan}  computes the ionization and temperature structure of the slab. We then use this structure to run {\sc noar} in order to simulate the effects of photon {energy} shifts due to Comptonization.
We run {\sc noar} twice: once illuminating the slab by the power law, and then invert the structure and illuminate
it with a black-body SED in order to simulate the soft flux from the disk below the corona.
We then normalize the obtained spectra from photons to physical unit by correlating them with {\sc titan}.
In this way, despite two independent runs of {\sc noar}, we can obtain the combined spectrum of reflection and
soft comptonization or either of the components separately.
In the end, the final spectra are analyzed to extract values shown in the maps.\\

We have done hundreds of simulations varying the 5 parameters $n_{\rm H}$, $\tau_{\rm cor}$, $q_{\rm h}$,  {\boldmath$\xi_0$} and $T_{\rm bb}$. 
For each simulation, we can compute the value of $\chi$ given by (see Eq. \ref{eqChiH} and \ref{eqHgen}):
\begin{equation}
\chi =\frac{\tau_{\rm cor}Q}{\frac{B}{4+3\tau_{\rm cor}}+\frac{\tau_{\rm cor}Q}{2}}\label{eqChisimu}
\end{equation}
Some of these simulations did not converge in some part of the parameter space. {The true reason is not clear yet.} The matter on the back side of the cloud can become too cool and too optically thick, below the actual limitations of the {\sc titan} code. {But it could also be linked to thermal instabilities that could exist at large $\tau$. While it is not a problem simple to understand and solve, this has no impact on the results} of the present paper since this part of the parameter space (mainly characterized by $q_{\rm h}<10^{-23}$ erg s$^{-1}$ cm$^{3}$) is not relevant for emission in the soft X-ray band (too low corona temperature). For this reason, we present only simulations with $q_{\rm h}>10^{-23}$ erg s$^{-1}$ cm$^{3}$.
We plan to improve {\sc titan} simulations for gas of much larger optical depth in our future work. 

Moreover, since the main goal of this analysis is to study the spectral properties of the warm corona, we only consider the intrinsic warm corona emission spectrum, without any reflected components due to the illumination from the hot corona. Reflection due to external illumination will be considered in a forthcoming paper.

\begin{figure*}
\begin{center}
\begin{tabular}{cc}
\includegraphics[height=0.49\textwidth]{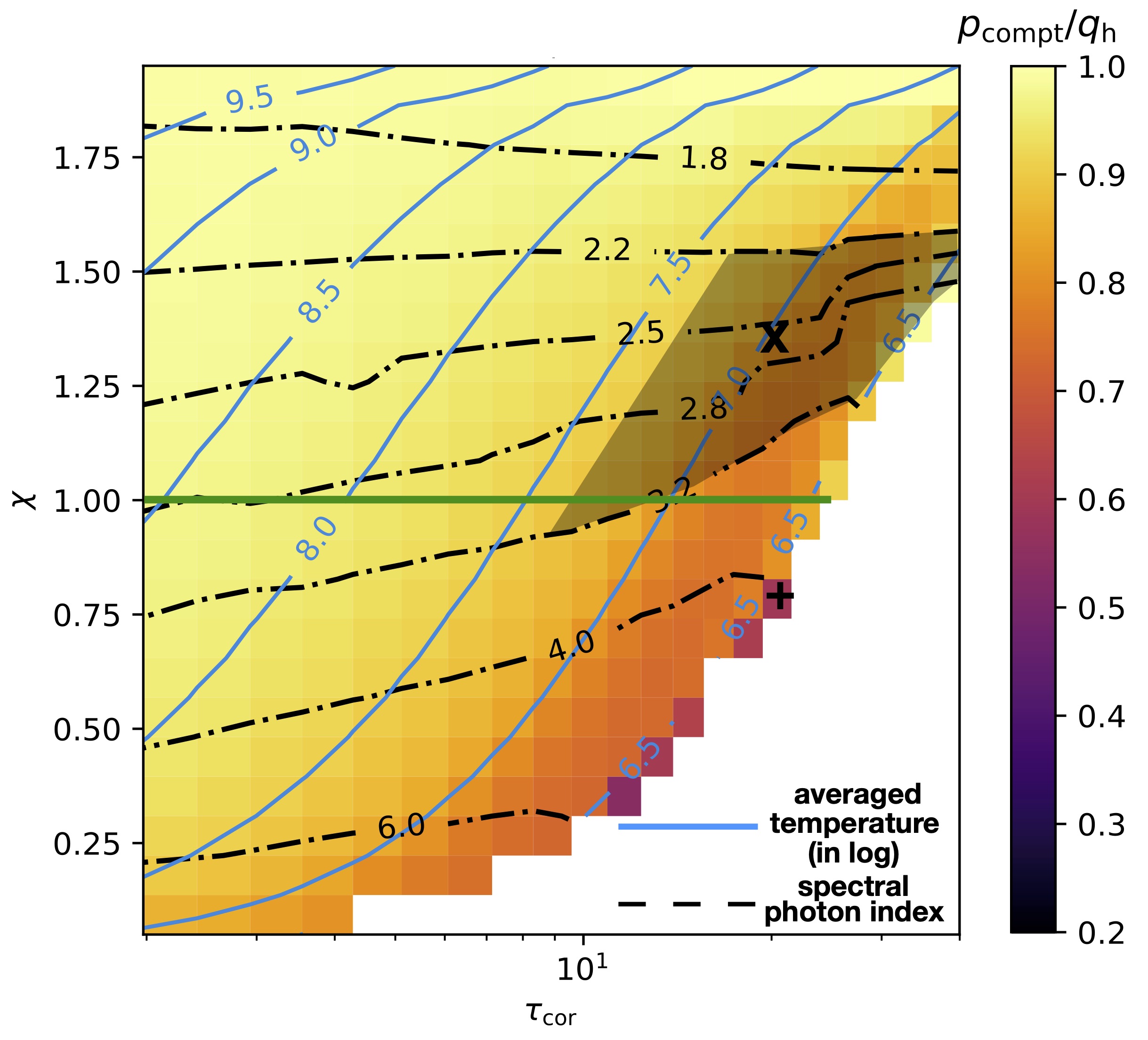} & \includegraphics[height=0.48\textwidth,width=0.4\textwidth]{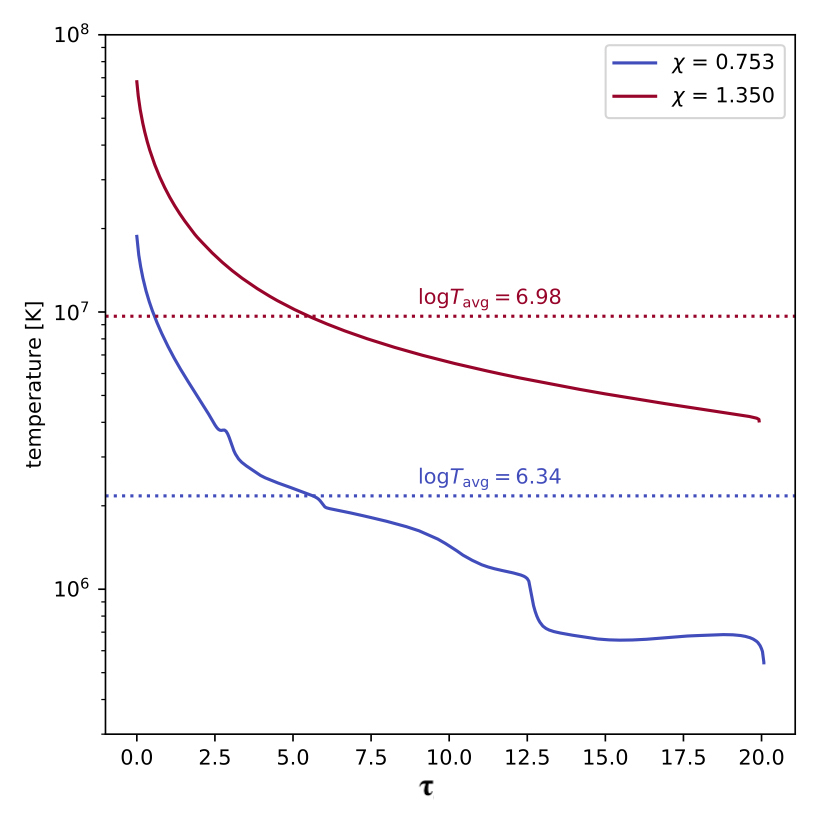}
\end{tabular}
 \caption{{\bf Left:}  Same map as in Fig. \ref{FigSimu1ab} but with the contours of the logarithm of the vertically average corona temperature (in Kelvin, blue solid line) and the contours of the photon indices of the output spectrum produced by the corona (black dot-dashed lines). The other parameters are fixed to $T_{\rm bb}=10^5$ K,  {\boldmath$\xi_0=10^3$} erg s$^{-1}$ cm, and $n_{\rm H}=10^{12}$ cm$^{-3}$. Similarly to Fig. \ref{FigSimu1ab} the green line corresponds to the condition $\chi=1$. The darkened area corresponds to the usual observational constraints of the temperature (0.1-2 keV) and photon index (2.2-3.2) of the soft X-ray excess. {\bf Right:} Vertical corona temperature profiles for $\tau_{\rm cor}=20$ and $\chi=1.35$ and $0.753$. The corresponding average temperatures $\displaystyle T_{\rm avg}=\frac{1}{\tau_{cor}}\int_{0}^{\tau_{\rm cor}} T(\tau)d\tau$ are indicated on each figures.}
\label{FigSimu2ab}
\end{center}
\end{figure*}

\subsection{Results}
We present in the two following sections the results from a subset of simulations where the warm corona density $n_{\rm H}=10^{12}$ cm$^{-3}$, the disk temperature $T_{\rm bb}=10^5$ K and the ionisation parameter  {\boldmath$\xi_0=10^3$} erg s$^{-1}$ cm. These are typical values \citep[e.g.][]{ree84} for AGN of $10^8$ solar masses black hole like e.g. Mkn 509. We discussed the effects of these parameters on our solutions in Sect. \ref{sectDep}.


\subsubsection{Compton dominated}
We have reported, in the left of Fig. \ref{FigSimu1ab}, the map of the ratio of the Compton cooling $p_{\rm compt}$ to the warm corona internal heating $q_{\rm h}$ in the $\chi$-$\tau_{\rm cor}$ plane. The color scale goes from 20\% to 100\%. Clearly for a very large part of the parameter space, Compton cooling is the dominant ($>$90\%) cooling process that balances the internal heating of the warm corona. This is the first important result of this work. It justifies the main hypothesis used in the two-coronae approach where the warm corona emission is modelled with pure Comptonisation codes.\\ 

The Compton cooling becomes less dominant at lower $\chi$ and larger $\tau_{\rm cor}$ where free-free and line processes start to play a role. We have reported in the right of Fig.  \ref{FigSimu1ab} the vertical distribution of the different heating and cooling processes for two simulations with $\tau_{\rm cor}=20$ but with two different values of $\chi=1.35$ and $0.753$. This corresponds to the two simulations indicated by a "x" and a "+" respectively on the left of Fig. \ref{FigSimu1ab}. The Compton cooling represents respectively more than 90\% and less than 70\% of the total cooling processes.\\ 

The solid green line overplotted in Fig. \ref{FigSimu1ab}-left corresponds to the condition $\chi=1$ i.e. the value expected in the case of radiative equilibrium between a slab corona covering entirely a non-dissipative disk (see Sect. \ref{transrad}).  The region above the green line agrees with a non-dissipative disk covered by a patchy corona. 
The dashed contours in Fig. \ref{FigSimu1ab}-left correspond to different values of the covering factor from 0.95 to 0.40 (see Eq. \ref{eqA} and \ref{eqChiA} for the general expression of $\chi$ in function of the covering factor). On the other hand, the region below the green lines agrees with a slab corona above a dissipative disk.\\


\subsubsection{Temperature and Spectral shape}
\begin{figure}[t]
\begin{center}
\includegraphics[width=\columnwidth]{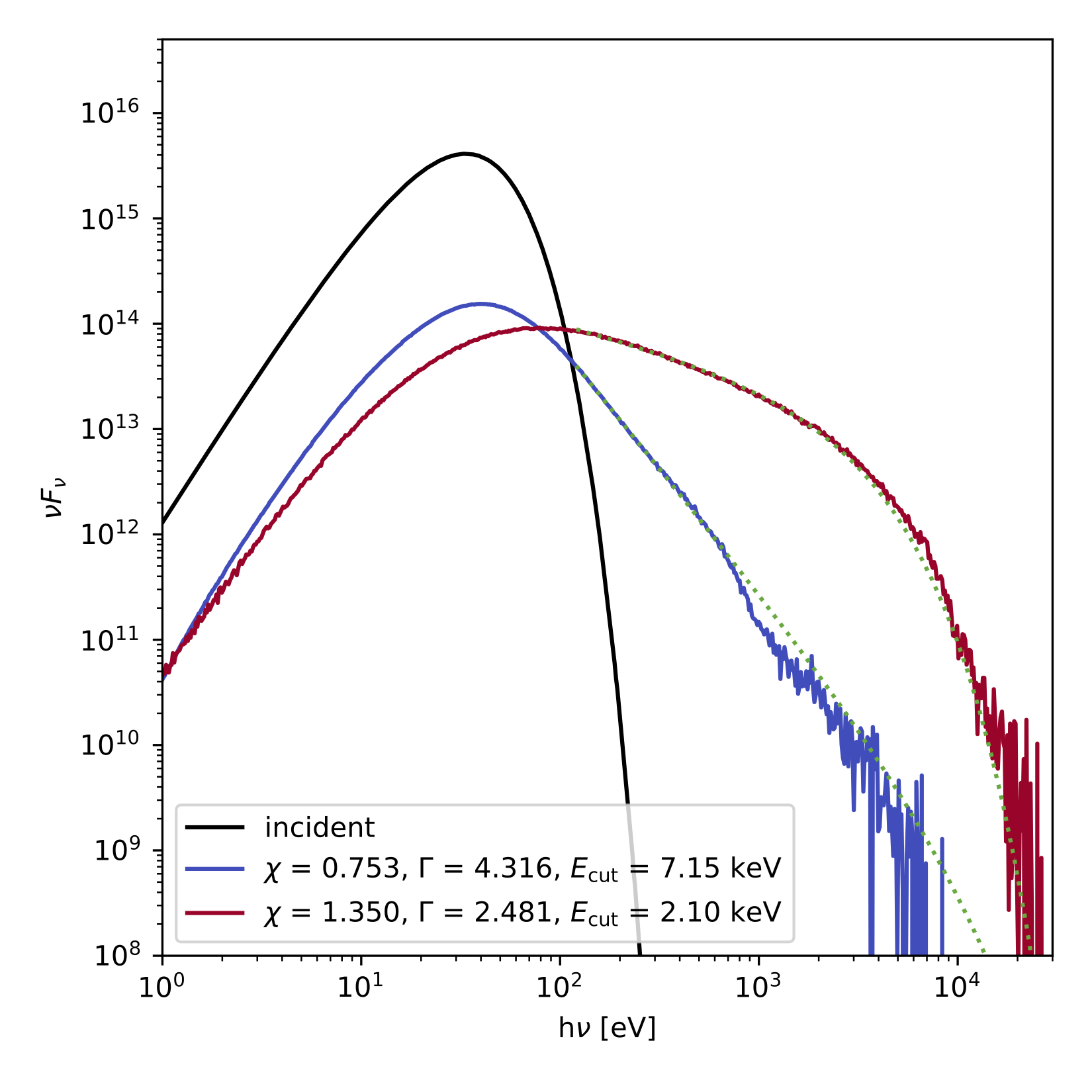}
 \caption{Two examples of simulated spectra emitted by the warm corona for $\tau_{\rm cor}=20$ and two different values of $\chi$=1.35 (red)  and $0.753$ (blue). The best fit values for the photon index and high energy cut-off when the spectra are fitted with a high energy exponential cut-off power law are indicated on the figure. The simulation corresponding to the red spectrum is represented by the "x" in Fig. \ref{FigSimu2ab} and is in the parameter space which agrees with the spectral shape of the soft X-ray excess. On the contrary,  the simulation corresponding to the blue spectrum is represented by the "+" in Fig. \ref{FigSimu2ab} and has a much steeper spectrum compared to the usual values of the soft X-ray excess. We have also reported in black the corresponding black-body spectrum emitted by the disk at the bottom of the warm corona.}
\label{FignuFnu}
\end{center}
\end{figure}
We have reported in Fig. \ref{FigSimu2ab}-left,  the same map of the ratio of the Compton cooling $p_{\rm compt}$ to the warm corona internal heating $q_{\rm h}$ as in Fig. \ref{FigSimu1ab} but we have now over plotted the contours of the average corona temperature in dotted lines.  This temperature is defined as $\displaystyle T_{\rm avg}=\frac{1}{\tau_{\rm cor}}\int_{0}^{\tau_{\rm cor}} T(\tau)d\tau$.  An example of the temperature vertical distribution $T(\tau)$ for $\tau_{\rm cor}=20$ and $\chi=1.35$ and 0.753 are plotted in Fig. \ref{FigSimu2ab}-right. The Compton cooling being dominant, the electron temperature naturally decreases with $\tau$ due to the increase of the soft photons flux while the heating $Q$ is assumed to be constant (see R15). And as expected, for larger $\chi$ (i.e. larger corona heating with respect to the total radiative cooling) the corona temperature is higher. \\

For each simulation, our code also computes the spectrum emitted upward by the warm corona. Each spectrum is fitted above 100 eV with a cut-off power law. The dot dashed lines reported in Fig. \ref{FigSimu2ab}-left  correspond to different spectral photon indices of these spectra between 1.8 to 4.5. As an example, we have reported in Fig. \ref{FignuFnu} the simulated spectra emitted by the warm corona for $\tau_{\rm cor}=20$ and $\chi=1.35$ (red line)  and 0.753 (blue line). In the former case, the spectrum can be fitted with a cut-off power law with $\Gamma\simeq$ 2.5 and $E_{\rm c}\simeq$ 2 keV while for $\chi=0.753$ we find $\Gamma\simeq$ 4.3 and $E_{\rm c}\simeq$ 10 keV\footnote{It is known that an exponential cut-off is a poor approximation of the cut-off of realistic comptonisation models (e.g. \citealt{pet00}). It overpredicts the corona temperature by a factor 2-3 which is indeed what we observe.}. 
We have also reported in this figure the black-body spectrum (black line) emitted by the disk at the bottom of the warm corona. The difference in flux between the disk black-body and the corona spectra results from the large opacity of the warm corona. It reduces the flux of the disk, after scattering through the corona, by a factor $\displaystyle\sim\frac{1}{1+\frac{3}{4}\tau_{\rm cor}}$.\\ 

Observationally, when fitted with a Comptonisation model, the photon index of the soft X-ray excess is generally between 2.2 and 3.2 and peaks around 2.5  and the temperature is between $\sim$0.1 and $\sim$2 keV (see e.g. P18 and references therein).  Very interestingly, there is a region  of our parameter space with the right range of temperatures and photon indices. It corresponds to the dark area in Fig. \ref{FigSimu2ab}-left. Moreover, this region is not far from the solid green line which corresponds to the condition $\chi=1$ i.e. the expected value for a slab corona covering entirely a non-dissipative disk. Actually, when we compare the position of the dark area of Fig. \ref{FigSimu2ab}-left with Fig. \ref{FigSimu1ab}-left, it is in better agreement with a slightly patchy corona with a covering factor in between $\sim$0.9 and $\sim$1. This is another important result of these simulations. This confirms the modelling done by R15 and agrees with the observational estimates of P13 and P18. \\

{As explained in Sect. \ref{titancode}, since {\sc titan} does not take into account the Compton heating of the radiation field, the plasma temperature can be over-estimated in our computations. However, this effect is only significant for $\chi>1$, and only the temperatures in the upper part of Fig. \ref{FigSimu2ab}-left are expected to be smaller than the ones computed by {\sc titan}. A precise estimates is not easy to do, but it means that the dark region is expected to be even larger than the one shown in Fig. \ref{FigSimu2ab}.}

It is also worth noting that  we cannot exclude disk-corona geometries that would combine simultaneously a lower covering factor and a more dissipative disk. 
While not impossible, the required fine tuning between the covering factor and the disk intrinsic dissipation to be in the right ranges of observed temperatures and spectral indices is not easy to explain from a physical point of view, making this option relatively unlikely.

\subsubsection{Dependences on  {\boldmath$\xi_0$}, $n_{\rm H}$ and $T_{\rm bb}$}
\label{sectDep}
\begin{figure*}[t]
\begin{center}
\includegraphics[width=\textwidth]{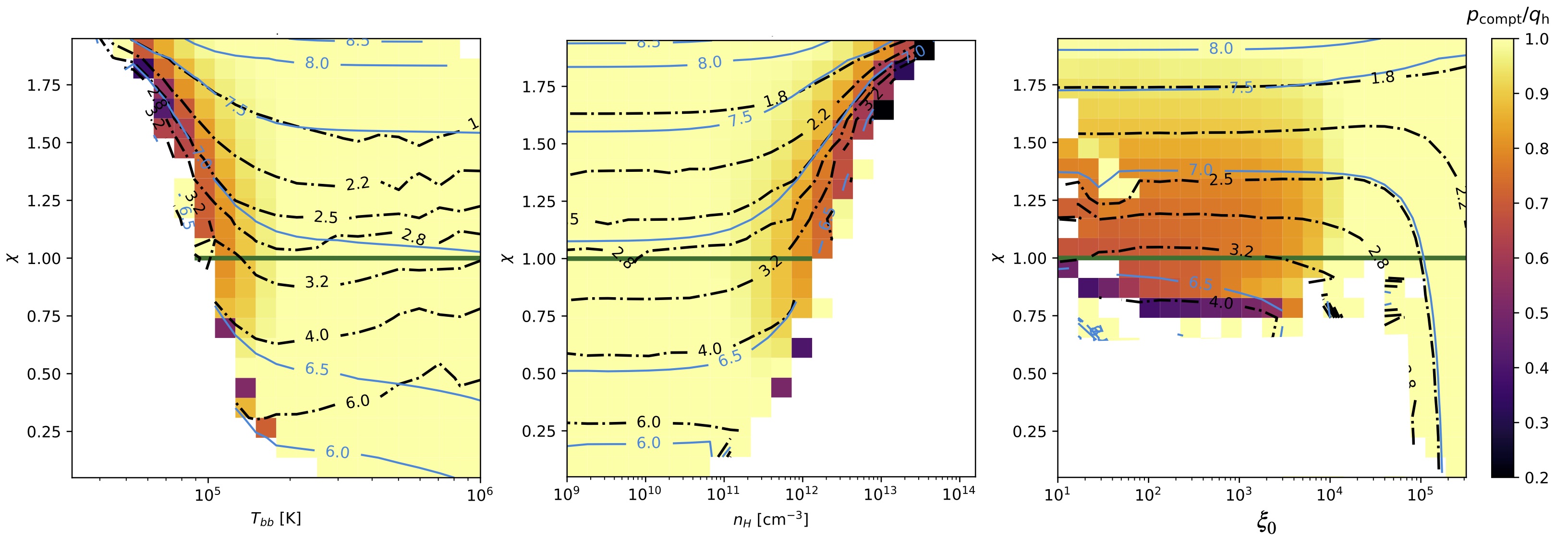}
 \caption{Contours of the warm corona temperature (blue lines) and the warm corona spectral photon index (black dashed lines) in the planes, from left to right, $\chi$-$T_{\rm bb}$, $\chi$-$n_{\rm H}$  and $\chi$-{\boldmath$\xi_0$}.  When $T_{\rm bb}$, $n_{\rm H}$ or  {\boldmath$\xi_0$} are not variable they are fixed to $T_{\rm bb}=10^5$ K, $n_{\rm H}=10^{12}$ cm$^{-3}$ and  {\boldmath$\xi_0=10^3$ erg s$^{-1}$} cm. The corona optical depth $\tau_{\rm cor}$ is fixed to 20. The background colors correspond to the Compton cooling to corona heating fraction whose color scale is reported on the right. Similarly to Fig. \ref{FigSimu1ab} the green line corresponds to the condition $\chi=1$.}
\label{FigXi}
\end{center}
\end{figure*}
In order to check the dependency of our results on the different parameters of the simulations, we have tested different disk temperatures $T_{\rm bb}$, different corona densities $n_{\rm H}$ and different values of the ionisation parameter  {\boldmath$\xi_0$} of the illumination from above. We have reported in Fig. \ref{FigXi}  the contours of the warm corona temperature (blue lines) and warm corona spectral photon index (black dashed lines) in the planes, from left to right, $\chi$-$T_{\rm bb}$, $\chi$-$n_{\rm H}$  and $\chi$-{\boldmath$\xi_0$}. When $T_{\rm bb}$, $n_{\rm H}$ and {\boldmath$\xi_0$} are not the variable parameter they are fixed to $10^5$ K, $10^{12}$ cm$^{-3}$ and $10^3$ erg s$^{-1}$ cm respectively. The corona optical depth $\tau_{\rm cor}$ is fixed to 20.\\

 These figures show that for large $T_{\rm bb}$ or low $n_{\rm H}$ or {\boldmath$\xi_0$}, the temperature and the spectral photon index of the warm corona emission are almost insensitive to these parameters. In the case shown here, i.e. $\tau_{\rm cor}=20$, this corresponds to $T_{\rm bb}>10^5$ K, $n_{\rm H}<10^{12}$ cm$^{-3}$ or  {\boldmath$\xi_0 < 10^4$}. In this region of the parameter space, the properties of the warm corona are mainly driven by $\chi$. For smaller $T_{\rm bb}$ or larger density  $n_{\rm H}$ or ionisation parameter  {\boldmath$\xi_0$} however, the radiative and spectral properties of the corona depend significantly on them. The background colors on Fig. \ref{FigXi}  correspond to the Compton cooling to corona heating fraction $p_{\rm compt}/q_{\rm h}$ whose colour scale is reported on the right of the figure. \\

The dependencies of the corona emission spectral shape on $T_{\rm bb}$ and $n_{\rm H}$ can be explained by the properties of the Comptonisation process in an environment in radiative equilibrium.  At low density or high disk temperature our simulations show that the corona emission is dominated by Comptonisation. In this case, as far as the heating over cooling of the warm corona keeps constant no significant spectral variability is expected. The corona heating is directly related to the sum of the internal heating $Q$ and the illumination  {\boldmath$\xi_0$} while the cooling is related to $B=\sigma T_{\rm bb}^4$. They are all related one with each other through the net flux crossing the base of the corona (see Eq. \ref{eqHgen}). As far as the illumination is negligible with respect to warm corona internal heating power, a constant $\chi$ implies a constant heating/cooling ratio and the warm corona temperature is not expected to change. The warm corona optical depth being fixed, the spectral shape of the warm corona emission stays roughly constant. This is very similar to the constancy of the spectral shape of the emission from a hot optically thin corona in radiative equilibrium with the surrounding soft photon field (e.g. \citealt{haa91,haa93,pet00,tor18}).\\

 For density larger than $\sim$10$^{12}$ cm$^{-3}$ and disk temperatures lower than $\sim$10$^5$ K however, the dependence on these parameters becomes important. This is due to the fact that, in this part of the parameter space, the corona radiative processes are less dominated by Comptonisation, either because the soft seed photons flux decreases (at low $T_{\rm bb}$) or the free-free and bound-free processes increase (at high density) through their dependence to the density squared. The previous discussion in terms of radiative equilibrium does not hold anymore. At constant $\chi$, decreasing $T_{\rm bb}$ or increasing $n_{\rm H}$ results in a decrease of the temperature and a steepening of the warm corona spectrum. These conditions are thus less favourable to produce spectra that could reach the soft X-ray band and produce a significant soft X-ray excess. \\

\begin{figure}[t]
\begin{center}
\includegraphics[width=\columnwidth]{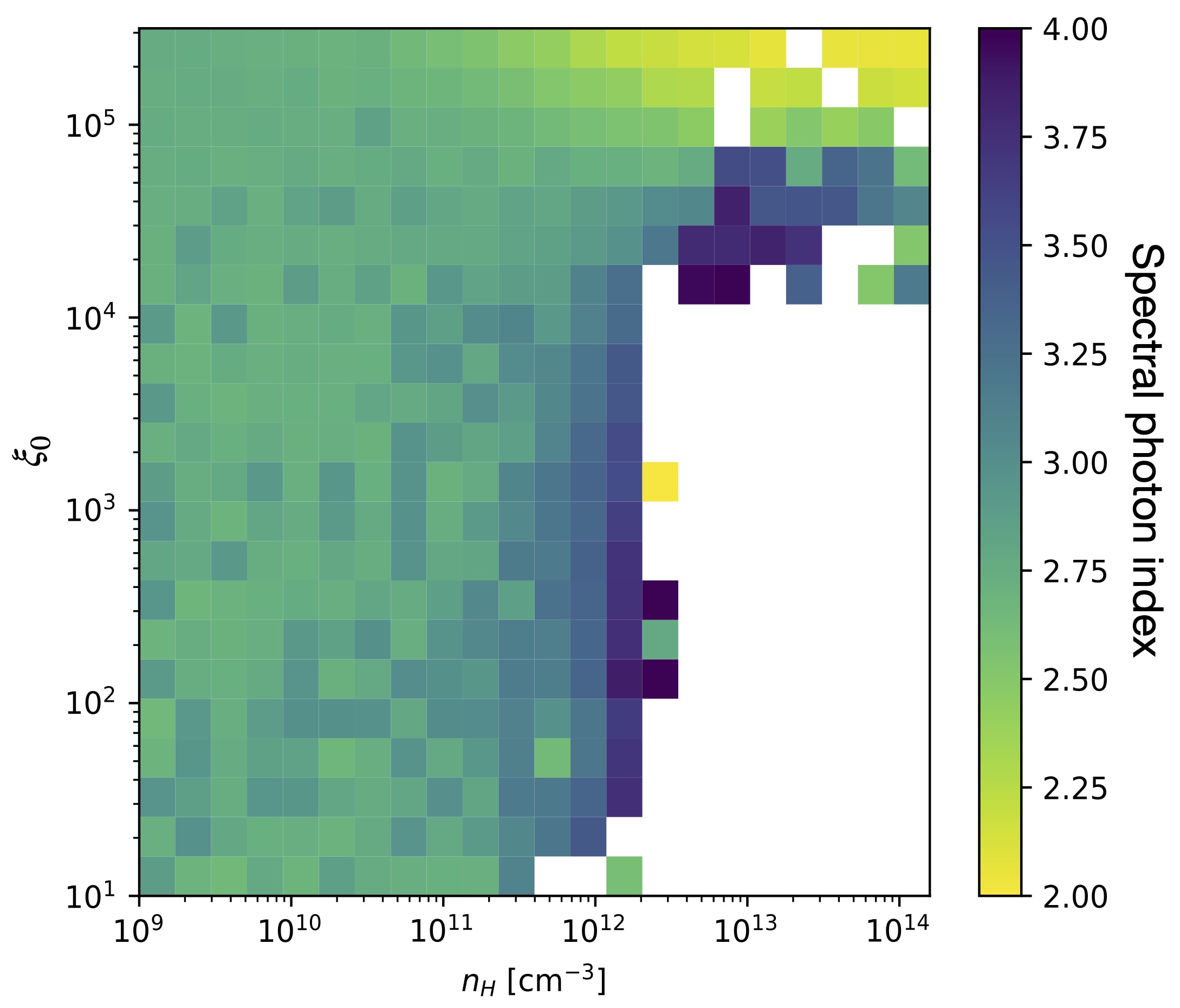} 
\caption{Map of the spectral photon index of the warm corona emission in the {\boldmath$\xi_0$}-$n_{\rm H}$ plane. The other parameters are fixed to $T_{\rm bb}=10^5$ K, $\chi=1$ and $\tau_{\rm cor}=20$.}
\label{Figxi-n}
\end{center}
\end{figure}
The dependence of the spectral shape and temperature of the warm corona on the ionisation parameter {\boldmath$\xi_0$} is quite different. They are almost independent on {\boldmath$\xi_0$} unless it becomes larger than a few 10$^4$ erg s$^{-1}$ cm. While, for such values of {\boldmath$\xi_0$}, the warm corona is dominated by Compton cooling, the spectral shape (photon index and temperature) varies with {\boldmath$\xi_0$} and the lower $\chi$ the larger the variation. Actually, increasing {\boldmath$\xi_0$} increases the heating of the corona which is not compensated by a similar increase of the cooling given the small fraction, due to the large corona optical depth, of the illumination flux that reaches and is reprocessed in the disk. At constant $\chi$ this implies an increase of the warm corona temperature and a hardening of the spectra. This is interesting because it shows that even for small $\chi$, the warm corona can produce spectra in agreement with the soft X-ray excess spectral properties if the illumination is strong enough. \\

Since the spectral shape and temperature of the warm corona vary in a similar way for opposite variations of the illumination and the density, an increase of the illumination could attenuate the effects produced on the spectral shape by an increase of the density. To verify this hypothesis, we have reported in Fig. \ref{Figxi-n} the map of the spectral photon index in the {\boldmath$\xi_0$}-$n_H$ plane. Simulations with large density ($>10^{14}$) can indeed produce spectra with the right spectral photon index (i.e. in the range [2.2-3.2]) for large ionisation parameter of the illumination ({\boldmath$\xi_0$} larger to a few $10^4$ erg s$^{-1}$ cm). In this region of the parameter space, the spectral photon index is also quite sensitive on {\boldmath$\xi_0$}.


\subsection{Emission/absorption lines}
\begin{figure}[t]
\begin{center}
\includegraphics[width=\columnwidth]{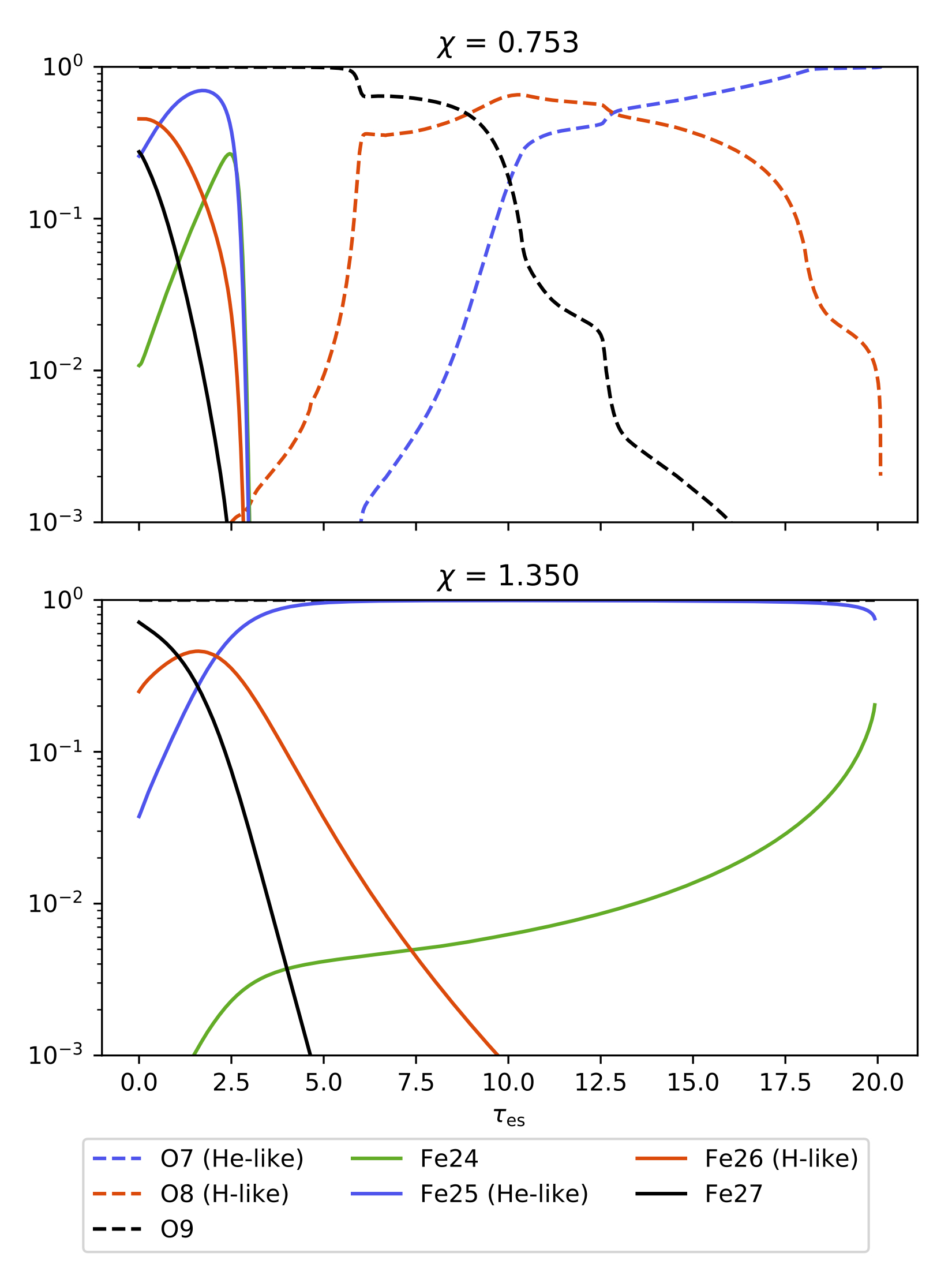}
 \caption{Ionisation state of Iron and Oxygen across the corona for $\tau_{cor}$=20 and $\chi$=0.753 (top) and 1.35 (bottom). The other parameters are fixed to $T_{\rm bb}=10^5$ K, {\boldmath$\xi_0=10^3$} erg s$^{-1}$ cm, and $n_{\rm H}=10^{12}$ cm$^{-3}$. Note that for chi=1.350 Oxygen is fully ionized throughout the entire corona. }
\label{FigIon}
\end{center}
\end{figure}

As said before, in the region of the parameter space which agrees with the observational constraint of the soft X-ray excess (dark area in Fig. \ref{FigSimu2ab}-left) Comptonization is the dominant cooling process. This is of course directly related to the assumed extra heating $Q$ which keeps the corona temperature high enough and strongly weakens the coolings due to free-free or line emission. This also means that the presence of emission/absorption features in the output spectrum is expected to be small. We have reported in Fig. \ref{FigIon} the ionisation fraction of iron and oxygen for $\tau_{\rm cor}=20$ and $\chi=0.753$ and $1.35$. A large fraction of iron is fully ionized at the corona surface ($\sim$ 30\% for $\chi$=0.753 and $>70$ \% for $\chi$=1.35). So ionized iron lines are potentially expected but around 7 keV and certainly not in the soft X-rays. Correspondingly, Oxygen is always fully ionized very deeply inside the corona (up to $\tau$=5 for $\chi$=0.753 but in the entire corona for $\chi$=1.35). There is no chance for line emission to get out of the corona without being completely smeared by Comptonization effects. This confirms that no lines are expected from this element and we have checked that this is the case for all the other elements especially those producing lines below 2 keV (e.g. Silicon, Magnesium).  In conclusion the spectra presented in Fig. \ref{FignuFnu} do not show any strong emission/absorption features in the energy range where soft X-ray excess is observed. 

\section{Discussion}
\label{summary}
\subsection{Brief summary}
We have used in this paper  the state of the art radiative transfer code {\sc titan} coupled with the Monte Carlo code {\sc noar} to simulate the physical and radiative properties of optically thick and dissipative coronae for different sets of optical depth, density and internal heating power. We also assumed illumination from above by X-rays, characterized by the ionisation parameter {\boldmath$\xi_0$}, and from below by an optically thick accretion disk radiating as a black-body of temperature $T_{\rm bb}$.  The simulations included all the relevant cooling and heating processes. The  vertical temperature profile of the corona was computed assuming radiative equilibrium and the total emitting spectrum (continuum and emitting/absorbing features) is produced.\\

We have shown that, in a large part of the parameter space, Compton scattering process is the dominant cooling process which balances the internal heating power of the corona. Moreover, there is a region of the parameter space where warm ($kT$ in 0.1-2 keV) and optically thick ($\tau\sim20$) coronae produce spectra in agreement with the observed soft X-ray excess spectral shape. No strong emission/absorption features are observed. This region of the parameter space is also consistent with a slightly patchy corona (covering factor between 0.9 and 1) in radiative equilibrium above a non-dissipative accretion disk, all the power being released in the warm corona.  These results are in very good agreement with the physical constraints obtained from the fit of the soft X-ray excess with warm Comptonisation model (e.g. P13, P18)  and with the theoretical estimates done by R15. Our more accurate and more physical simulations differ however significantly from the ones done by \cite{gar18}, where some fundamental ingredients (like the corona internal heating power) were not included. Our results show that a warm and optically thick corona can indeed explain the
soft X-ray excess, provided that such coronal heating is allowed.\\

\subsection{An accretion disk and two powerful coronae to explain the broad band spectral emission of radio-quiet AGNs}
The results of our analysis crucially depend on the assumption that internal heating power is released in the warm corona up to large optical depth ($\tau>$10). Indeed, if $Q$ goes to 0 the equations shown in Sect. \ref{transrad} reduce to the standard radiative profiles of grey atmospheres. On the other hand, the radiative properties of the warm corona are directly related to the conditions of radiative equilibrium with the accretion disk. A powerful warm corona covering almost entirely a non-dissipative disk will naturally produce spectra in agreement with the observed soft X-ray excess spectral properties.\\

 This is very similar to the case of the hot corona that produces the hard X-ray (above $\sim$2 keV) emission in radio-quiet AGNs. This hot corona is also believed to be dominated by Comptonisation and to have its own local heating (e.g. \citealt{mer01}). And its radiative properties are strongly constrained by its radiative equilibrium with the surrounding colder regions, sources of soft seed photons (e.g. \citealt{haa91,haa93b,haa97}). In order to agree with hard X-ray observations however (spectral photon index $\Gamma\sim$1.9, temperature $kT\sim$100-300 keV) the hot corona has to be optically thin and significantly patchy (see also discussion in P18). \\
 
Combining these constraints for the hot and warm coronae, the UV/soft X-ray/hard X-ray spectral emission of radio quiet AGNs would result from the radiative equilibrium between 3 main components: a weakly-dissipative accretion disk, and two powerful coronae, one optically thick (and warm) and one optically thin (and hot) (see also e.g. \citealt{kub18}). Variations in the geometry and/or in the repartition of the heating power among these three components would produce the palette of observed broad band spectral shapes.

\subsection{Origin of the warm corona heating}
The immediate question raised by this work is the origin of the warm corona internal heating power and the reason why it should be most preferentially released in the upper layer of the accretion flow rather than in its deeper layers as assumed in the standard accretion disk theory (e.g. \citealt{sha73}). While a detailed answer to this question is beyond the scope of this paper, we note that recent
simulations and modelling of accretion flows seem to bring some support to it. Clearly, the magnetic field plays a crucial role here and we naively expect that the buoyant rise of magnetic field generated within magnetically dominated accretion disk could transport substantial amounts of energy vertically, and then heat up the upper layers of the accretion flow  (e.g. \citealt{mer00,hir06,beg15}).  Very recently, through radiative transfer computation in hydrostatic and radiative equilibrium, \cite{gron19} have shown that gas heating by magnetic dynamo (following \citealt{beg15}) is indeed able to heat the upper layers of the accretion flow up to a few keV for optical depths as large as $\sim$ 10.  By construction, in this approach, the underlying disk cannot be completely passive but these results give a nice support for the presence of such local heating close to the disk surface.\\
Recent general-relativistic magnetohydrodynamic (GRMHD) simulations also indicate that magnetized disks accrete preferentially through the disk atmosphere where the disk is magnetically dominated \citep{beck09,zhu18,mish19,jia19}.  
Such "coronal accretion" could play the role of our warm corona if some heating, linked to the accretion phenomena, could be locally released.
In another domain of application, 3D radiation-MHD calculations, when applied to protostellar disks, show that the disk surface layer can indeed be heated by dissipating the disk magnetic field, while the disk interior is ``magnetically dead'' and colder than a standard accretion disk (e.g. \citealt{hir11}). 

These different results suggest that the presence of local heating power at the disk surface and up to a few optical depth is physically plausible.\\ 

\subsection{Potential impacts of the presence of a warm corona in the upper layers of the accretion flow}
The presence of a warm plasma in the upper layers of the accretion flow could have several effects on the dynamical, spectral and temporal properties of the accretion flow itself. It is already known that a warm corona can stabilize the underlying accretion disk, otherwise subject to radiation pressure instability (see \citealt{cze19} for a very recent review). More unexpectedly, the presence of a warm corona in the quasar spectral energy distribution have been recently shown to be an important element to reproduce the extreme part of the ``main sequence''  of quasars\footnote{see \cite{mar18} for a recent review on the quasar main sequence}, especially quasars with particularly strong FeII emission \citep{pan19}. These two examples show already the wide variety of impacts expected from the existence of such a warm corona in the inner regions of AGN. We indicate here below a few other effects suggested by the present analysis.

\subsubsection{Outflows}
{A warm corona could play a major role in the production of disk outflows. The fact that the corona could be out of hydrostatic equilibrium was already pointed out by R15  and outflows may naturally appear (see also the discussion in \citealt{gron19}). Now, a precise understanding of these effects would require a numerical resolution of the disk/corona vertical equilibrium {in the gravitational potential of the central compact object} which is far out of the scope of the present paper where we simply assume a constant corona density. Saying that, we want to make here a few, admittedly qualitative, remarks.}

Interestingly, comparison of wind signatures in AGN and XrB with outflow MHD simulations suggest high ejection efficiencies $p>0.5$ with $\dot{M}_{\rm acc}\propto R^{\rm p}$ \citep{zha13,fuk15,cha16,fuk18}. And such high values of $p$ are more easily reached if there is some heating deposit at the disk surface (e.g. \citealt{cas00b,fer04}). Since, as we have shown in this study, the presence of a warm corona could also explain the soft X-ray excess observed in a large number of AGNs, we could expect to find stronger wind signatures in objects where a strong soft X-ray excess is also present. To our knowledge, no such relation has been observed yet. The observational signatures are not obvious however given the variability of the wind signatures and the potential dependence on physical parameters like e.g. the disk/corona inclination. 
As we have seen, the importance and the spectral shape of the warm corona emission also depend on parameters like the disk temperature, the density of the warm corona or the illumination by the hot corona. So, even if there is significant heating deposit at the disk surface, which would favor disk outflows, the warm corona emission may not be always sufficient to produce any soft X-ray excess.

\subsubsection{Spectral variability}
We have seen in Sect. \ref{sectDep} that at low density, high disk temperature or low illumination the corona thermal properties and spectral emission depend almost exclusively on $\chi$. In consequence, no significant spectral variability is expected unless the warm corona/disk structure changes, e.g. the corona becomes more patchy ($\chi$ will increase) or the disk becomes more dissipative ($\chi$ will decrease). \\
For densities larger than $\sim10^{12}$ cm$^{-3}$, disk temperatures lower than $\sim10^5$ K or corona illumination with {\boldmath$\xi_0$} larger than $10^4$ erg cm s$^{-1}$ however, the dependence on these parameters becomes important. Moreover, the ranges of density and disk temperature in agreement with a soft X-ray excess spectral shape become narrower. Even if $\chi$ stays constant, small changes of $n_{\rm H}$, $T_{\rm bb}$ or  {\boldmath$\xi_0$} are expected to produce significant spectral variability of the warm corona emission and consequently on the soft X-ray range. Interestingly, high illumination can compensate the limitations due to too large densities or too low disk temperatures. But to obtain the right spectral shape  could require some fine tuning between these parameters.

\subsubsection{Reverberation lags}
Delays (so-called lags) between the different energy bands of the X-ray continua have been commonly detected in the past 10 years (e.g. \citealt{demar13,demar15,kar16,epi16,demar17}). On long time-scales hard X-rays lag the soft X-rays while at short time-scales the reverse is observed. The former are generally interpreted as intrinsic accretion fluctuations which drive the variability from the outer (and colder) parts of the accretion flow to the inner (and hotter) parts. The negative lags, at short time scale, are interpreted as reverberation lags due to the time delays between the changes in the direct X-ray continuum and the reprocessed, reflected X-rays from the disc (see \citealt{utt14,demar19} for recent reviews). Simulations to reproduce these lags generally assume the presence of a point-like X-ray source (the so-called lampost model) above a cold standard accretion disk (\citealt{emm14,cac14,cha15,epi16,chai16} 
but see \citealt{chai19,mah18,mah19} for radially stratified geometries). 
If this interpretation of the soft lags is correct, part of the soft X-ray emission should be explained by reflected photons.  Now, the presence of a warm corona at the surface of the accretion disk could modify this picture. First, the shape of the reflection component can be different if the presence of a warm corona at the surface of the disk is taken into account {since it can  significantly modify the ionisation conditions at the surface of the accretion flow}. Secondly, the soft X-ray photons emitted by the warm corona that will reach the observer will have a pathway clearly distinct from the hard X-ray photons. We can expect that a radially stratified geometry with the hot corona in the inner region and the warm corona outside, covering the inner part of the optically thick accretion disk, will produce time lags between the hard and soft energy ranges. The timing properties of a geometry of this type have been recently studied and applied to X-ray binaries \citep{mah18,mah19} and the frequency dependent lags between the X-ray bands have been successfully reproduced. {While these results strongly depend upon the assumed geometry of the accretion flow, which is basically unknown, they are quite encouraging and provide hopes for an application in AGN}.\\

%

\section{Conclusion}
We have studied the radiative equilibrium and spectral emission of an optically thick plasma (the so-called warm corona) heated by an internal source of power and illuminated from above by a hard X-ray power law and from below by a disk black-body. Such a plasma has been proposed to explain the soft X-ray excess observed in a large number of AGNs, however its radiative properties have not been studied very much yet. 
We have made simulations using the last version of the radiative transfer code {\sc titan}  coupled with the  Monte-Carlo code {\sc noar}, the latter fully accounting for Compton scattering of continuum and lines. We have tested different plasma optical depths, densities, internal heating powers as well as disk temperatures and ionisation parameters of the power law illumination. The main conclusions are the following:
\begin{itemize}
\item In a large part of the parameter space, the Compton cooling is dominant across all the corona. For a given optical depth, higher internal heating, higher disk temperatures, lower densities or higher illumination favoured this dominance. 
\item The emission and absorption lines are also expected to be weak or completely absent due to the high ionisation state of matter in the upper part ($\tau$ of a few) of the corona where the ions are highly ionized and even completely stripped.
\item There is a sub-part of the parameter space where the spectral emission of the warm corona has spectral properties (spectral index, temperature) similar to those observed for the soft X-ray excess.
\item  This sub-part of the parameter space is consistent with a warm corona covering a large part of a weakly dissipative accretion disk. 
\end{itemize}
These results plainly confirm the modelling done by \citep{roz15} and agree with the observational estimates of \cite{pet13,pet18}. They rule out the limitations generally put forward (i.e. no Compton dominance, emission/absorption lines expected in the outgoing spectrum) concerning the warm Comptonisation modelling of the soft X-ray excess.\\

These conclusions rely however on the existence of internal heating power in the corona. This is a mandatory condition for the existence of such warm plasma above the accretion disk. The origin of such an internal heating power is unknown even if recent simulations and modelling of magnetized accretion flows seem to bring some support to it. 

The dependences of the warm corona properties with the model parameters (like e.g. the densities, the hard power law illumination) could explain the diversity of the soft X-ray excess spectral shape and even its absence in some AGNs. This will be studied in a forthcoming paper.



\paragraph{Note:}
During the completion of our paper, we became aware of a
work on a similar subject by \cite{bal20} (B20 hereafter). This paper is quite complementary to ours and we discuss here the main differences and similarities. First, in B20 the reflection is included in the output spectra. This makes the comparison to our spectra difficult. Similar to our work however, B20 generally assumes a constant corona densities.  But he considers a small range of corona densities of a few $10^{14}$ cm$^{-3}$, two orders of magnitudes larger than the typical values of $10^{12}$ cm$^{-3}$ used in our simulations. At such large densities, and even if we have difficulties to obtained TITAN solutions able to converge, we do not expect a spectral emission in agreement with the soft X-ray excess spectral shape (see Fig. \ref{FigXi} middle plot). The role of the illumination, above the warm corona, can certainly play a crucial role here as explained in Sect. \ref{sectDep} and in agreement with the conclusions of B20. But we do not simulate the same range of parameters so again a precise comparison is not easy to do.  Our study covers a large range of densities but lower than $10^{14}$ cm$^{-3}$. And we show that smaller  densities help to obtain better conditions to produce a warm corona with the required properties. Interestingly, B20 us also able to study the case of a warm corona with $\tau_{\rm cor}=10$ in hydrostatic equilibrium. But contrarily to R15, this hydrostatic equilibrium does not include with the presence of the underlying disk. Even so, the vertical decrease in density of the B20 hydrostatic solution favours the formation of the warm corona. Another important difference between B20 and our analysis is that the corona internal heating in B20 is a fraction of the total dissipation. In this case, increasing the coronal heating implies a decrease of the disk flux while in our calculations the two quantities are disconnected. This allows us to use the internal heating as a free parameter and to explore a larger range of values. In conclusion, the approaches of the problem between B20 and us are a bit different and our parameter spaces are not the same. We both conclude, however, on the possibility to have a warm corona with the correct properties to reproduce the observed soft X-ray excess of type 1 AGN.

\section*{Acknowledgments}
We thank the anonymous
referee for comments which helped us improve the clarity of the
paper. We warmly thanks Anne-Marie Dumont, who was providing us 
with the newest version of NOAR and the associated documentation. POP acknowledges financial support from CNES and the French PNHE. SB, ADR and GM acknowledges financial support from the Italian Space Agency under grant ASI-INAF 2017-14-H.O.
AR and DG acknowledge financial support from Polish National Science Center grants No.
 2015/17/B/ST9/03422 and 2015/18/M/ST9/00541. BC acknowledges the financial support by the National Science
Centre, Poland, grant No. 2017/26/A/ST9/00756 (Maestro 9). FU acknowledges financial support from ASI and INAF under INTEGRAL ``accordo ASI/INAF 2013-025-R1''.


\section{Appendix}
The general radiative equilibrium of a corona above a disk has already been discussed in P18. However a few mistakes are present in this paper. We show the correct equations here below in the light of the general results discussed in Sect. \ref{transrad}.\\ 

Following the notation of this paper, $F_{tot, disk}=\pi B$ is the total flux emitted by the disk that enters the corona, and $F_{cor}=4\pi Q\tau_{\rm cor}$ is the flux produced by dissipation inside the corona. In the case of a corona dominated by Comptonisation (i.e. zero absorption) in plan parallel geometry and coherent scattering (no photon energy gain during scattering) the transmitted and reflected back fluxes of the disk can be written:
\begin{eqnarray}
F_{disk,trans} &=& \frac{F_{tot, disk}}{1+\frac{3}{4}\tau_{cor}}=\frac{\pi B}{1+\frac{3}{4}\tau_{cor}}\\ 
F_{disk,ref} &=& \frac{\frac{3}{4}\tau}{1+\frac{3}{4}\tau}F_{tot, disk}=\frac{\frac{3}{4}\tau}{1+\frac{3}{4}\tau}\pi B
\end{eqnarray}
The observed flux emitted at the top of the corona is equal to:
\begin{equation}
F_{out}=\frac{F_{tot, disk}}{1+\frac{3}{4}\tau_{cor}}+\frac{F_{cor}}{2}=\frac{\pi B}{1+\frac{3}{4}\tau_{cor}}+\frac{4\pi Q\tau_{\rm cor}}{2}.\label{eqLobs}
\end{equation}
This equation assumes that $F_{cor}$ is distributed homogeneously in the warm corona and is entirely radiated isotropically i.e. that half of it is emitted upward while the other half is emitted backward.
\begin{figure*}[t]
\begin{center}
\includegraphics[width=\textwidth]{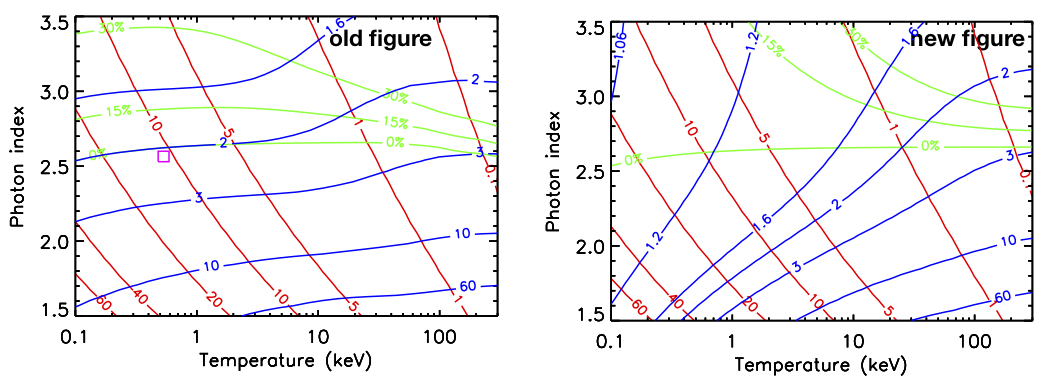}
 \caption{Left: figure published in P18. Right: new figure.}
\label{FigOldvsNew}
\end{center}
\end{figure*}

On the other hand, at the transition between the corona and the disk, the radiative equilibrium imposes:
\begin{eqnarray}
F_{tot,disk} &=& C_f (F_{rep} +F_{disk,intr})\\
	&=& C_f\left (\frac{\frac{3}{4}\tau_{cor}}{1+\frac{3}{4}\tau_{cor}}F_{tot,disk}+\frac{F_{cor}}{2}+F_{disk,intr}\right )\label{eqdisk}
\end{eqnarray}
where $F_{rep}$ is the  luminosity coming from the warm corona that is reprocessed in the disk (we assume an albedo of 0), $F_{disk,intr}$ is the instrinsic disk heating flux, and $C_f$ the warm corona covering factor.  A covering factor $C_f=1$ corresponds to a corona covering entirely the disk, while $C_f<1$ correspond to a patchy corona where part of the disk emission is radiated away without crossing the warm corona.

Introducing the amplification factor $A=1+F_{cor}/F_{tot, disk}$, these equations can be combined to give:
\begin{eqnarray}
A -1 = \frac{F_{cor}}{F_{tot, disk}} &=& 2\left ( \frac{1+\frac{3}{4}\tau_{cor}(1-C_f)}{C_f(1+\frac{3}{4}\tau_{cor})}\right ) -2\frac{F_{disk,intr}}{F_{tot, disk}}\label{eqA}\\
\frac{F_{disk,intr}}{F_{tot, disk}} &>& \frac{1}{1+\frac{3}{4}\tau_{cor}}-\frac{F_{cor}}{2F_{tot, disk}} =\left . {\frac{F_{disk,intr}}{F_{tot, disk}}}\right |_{min}\label{eqLint}
\end{eqnarray}
In the case $C_f$=1 and no intrinsik disk emission Eq. \ref{eqA} simply gives:
\begin{equation}
\frac{F_{cor}}{F_{tot, disk}} = \frac{2}{1+\frac{3}{4}\tau_{cor}}\label{eqAb}
\end{equation}
Moreover the amplification factor is related to the parameter $\chi$ defined in Eq. \ref{eqChi} via the following equation:
\begin{equation}
\chi=\frac{2(1+3\tau_{cor}/4)(A-1)}{2+(1+3\tau_{cor}/4)(A-1)}\label{eqChiA}
\end{equation}

Equations \ref{eqA} and \ref{eqLint} are different from Eq. (19) and (21) of P18 (assuming the albedo $a=0$ and that the corona covers entirely the disk, i.e.  $g=1$). While Eq. (19) and (21) of P18 agree with an optical depth of the order unity, they do not stand for large optical depth like the one expected in the warm corona. The main impact of the use of Eq.  \ref{eqA} and \ref{eqLint} of the present paper is on the expression of the amplification factor. It also modifies Fig. 1 of P18. We have reported in Fig. \ref{FigOldvsNew} the correct version of this figure. The conclusions of P18 still apply i.e. the observational characteristic of the soft X-ray excess (i.e. a photon index of $\sim$ 2.5 and a temperature in 0.1-2 keV) agree with an extended warm corona covering the disk which is mainly non-dissipative. \\


\end{document}